\documentclass[journal]{IEEEtran}
\usepackage{amsfonts}
\usepackage{slashbox}
\usepackage{amsmath}
\usepackage{multirow}
\usepackage{graphicx}
\usepackage{amsfonts}
\usepackage{amsmath,epsfig}
\usepackage{mathbbold}
\usepackage{stfloats}
\usepackage{amssymb}
\usepackage{url}
\usepackage{bm}
\usepackage{cite}
\usepackage{amsmath}
\usepackage{amssymb}
\usepackage{latexsym}
\usepackage{multirow}
\usepackage{epsfig}
\usepackage{graphics}
\usepackage{mathrsfs}
\usepackage{algorithmic}
\usepackage{algorithm}
\usepackage{subfigure}
\usepackage{slashbox}
\usepackage[all]{xy}

\usepackage{pifont}
\usepackage{bbding}
\usepackage{stmaryrd}
\usepackage{amssymb}
\usepackage{amsfonts}
\usepackage{epic}
\usepackage{stfloats}
\usepackage{latexsym}
\usepackage{epstopdf}
\usepackage{epic}
\usepackage{bm}
\usepackage{xcolor}
\usepackage{mathrsfs}
\usepackage{pifont}
\usepackage{bbding}
\usepackage{amsmath,epsfig}
\usepackage{mathbbold}
\usepackage{stmaryrd}
\usepackage{amssymb}
\usepackage{amsfonts}
\usepackage{epic}
\usepackage{graphicx}
\usepackage{subfigure}
\usepackage{enumerate}
\usepackage{stfloats}
\usepackage{latexsym}
\usepackage{epstopdf}
\usepackage{epic}
\usepackage{multirow}
\usepackage{stfloats}
\usepackage{bm}

\usepackage{booktabs}
\usepackage{color}

\newcommand{\q}{\mathbf q}
\newcommand{\Q}{\mathbf Q}
\newcommand{\A}{\mathbf A}
\newcommand{\w}{\mathbf w}
\newcommand{\pow}{\mathbf P}
\newcommand{\myfont}{\fontsize{8.5pt}{\baselineskip}\selectfont}

\begin{document}

\title{Joint Trajectory and Communication Design for Multi-UAV Enabled Wireless Networks }
\author{\IEEEauthorblockN{Qingqing Wu,  \emph{Member, IEEE}, Yong Zeng,  \emph{Member, IEEE}, and Rui Zhang, \emph{Fellow, IEEE}
\thanks{ The authors are with the Department of Electrical and Computer Engineering, National University of Singapore, email:\{elewuqq, elezeng, elezhang\}@nus.edu.sg. This work was presented in part at IEEE GLOBECOM 2017 \cite{wu2017joint}. This work was supported by the National University of Singapore under Research Grant R-263-000-B62-112. }}  }

\maketitle

\begin{abstract}
Unmanned aerial vehicles (UAVs) have attracted significant interest recently in assisting  wireless communication due to their high maneuverability, flexible deployment, and low cost.  This paper considers a multi-UAV enabled wireless communication system, where multiple UAV-mounted  aerial base stations (BSs) are employed to serve a group of users on the ground.  To achieve fair performance among users, we maximize the minimum throughput over all ground users  in the downlink communication by optimizing the multiuser communication scheduling and association jointly with the  UAVs' trajectory and power control. The formulated problem is  a mixed integer non-convex optimization problem that is challenging  to solve.  As such, we propose an efficient iterative algorithm for solving it by applying the block coordinate descent and successive convex optimization techniques. Specifically, the user scheduling and association, UAV trajectory, and transmit power are alternately optimized in each iteration.  In particular, for the non-convex  UAV trajectory and transmit power optimization problems, two approximate convex optimization problems are solved, respectively. We further show that  the proposed algorithm  is guaranteed to converge. To speed up the algorithm convergence and achieve good throughput, a low-complexity and systematic initialization scheme is also proposed for the UAV trajectory design based on the simple circular trajectory and the circle packing scheme. Extensive simulation results are provided to demonstrate the significant throughput gains of the proposed design as compared to other benchmark schemes.

\end{abstract}

\section{Introduction}
Unmanned aerial vehicles (UAVs), also commonly known as drones, have attracted significant attention in the past decade for various applications, such as surveillance and monitoring, aerial imaging, cargo delivery, etc \cite{valavanis2014handbook}.
As reported in \cite{globalUAVmarket}, the global market for commercial UAV applications, estimated at about 2 billion dollars in 2016, will skyrocket to as much as 127 billion dollars by 2020.
Equipped with advanced transceivers and batteries, UAVs are gaining increasing popularity in information technology (IT) applications  due to their high
maneuverability and flexibility for on-demand deployment. In particular, UAVs typically have high possibilities of line-of-sight (LoS) air-to-ground communication links, which are appealing to the wireless service providers \cite{lin2017sky}. To capitalize on this growing opportunity,  several leading IT companies have launched pilot projects,  such as Project Aquila by Facebook \cite{fb_UAV} and Project Loon by Google \cite{gl_UAV},   for providing ubiquitous internet access worldwide by leveraging the UAV/drone technology. The 3rd Generation Partnership Project (3GPP) is also looking up into the sky and studying aerial vehicles supported by Long Term Evolution (LTE) where the initial focus is on UAV \cite{3gpp}. In fact, with the approval of Federal Aviation Administration (FAA), Qualcomm and AT\&T have optimized  LTE networks for UAV communications  \cite{qualcom_UAV}, which aims to pave the way to a wide-scale deployment of UAVs in the upcoming fifth generation (5G) wireless networks, especially for mission-critical use cases.
 Meanwhile, extensive research efforts from the academia have also been devoted to employing UAVs as different types of
   wireless communication platforms  \cite{zeng2016wireless}, such as aerial mobile base stations (BSs) \cite{mozaffari2016unmanned,mozaffari2016efficient,al2014optimal,lyu2016placement,bor2016efficient}, mobile relays \cite{zhan2011wireless,zeng2016throughput}, and flying computing cloudlets \cite{loke2015internet,jeong2016mobile}. In particular,  employing UAVs as aerial BSs is envisioned as a promising solution to enhance the performance of the existing cellular systems. Depending on whether the UAV's high mobility is exploited or not, two different lines of research can be identified in the literature, namely static-UAV or mobile-UAV enabled wireless networks.

 The research on the static-UAV enabled wireless networks mainly focuses on the UAV deployment/placement optimization \cite{mozaffari2016unmanned,al2014optimal,mozaffari2016efficient,lyu2016placement,bor2016efficient}, with the UAVs serving as aerial quasi-static BSs to support ground users in a given area  from a certain altitude. 
As such, the altitude and the horizontal location of the UAV can be either separately or jointly optimized for different quality-of-sevice (QoS) requirements. In particular, the authors in \cite{al2014optimal} provide an analytical approach to optimize the altitude of a UAV for providing maximum coverage for ground users.
In contrast, by fixing the altitude, the horizontal positions of UAVs are optimized  in \cite{lyu2016placement} to minimize the number of required UAV BSs  to cover a given set of ground users.  In three-dimensional (3D) space, a drone-enabled small cell placement optimization problem is investigated in \cite{bor2016efficient} to maximize the number of users that can be covered.



Besides the UAV placement optimization,  exploiting the UAV's high mobility in the mobile-UAV enabled wireless networks is anticipated to unlock the full potential of UAV-ground communications. With the fully controllable UAV mobility,  the communication distance between the UAV and ground users can be significantly shortened by proper UAV trajectory design and user scheduling. This is analogous and yet in sharp contrast to the existing small-cell technology  \cite{wu2016energy,wu2016overview,zhang2016fundamental,wang2017joint}, where the cell radius is  reduced by increasing the number of small-cell BSs deployed, but at the cost of increased infrastructure expenditure. Motivated by this, the UAV trajectory design is rigorously studied in \cite{zeng2016throughput} and \cite{zeng2016energy} for a mobile relaying system and point-to-point energy-efficient system, respectively, where sequential convex optimization techniques are applied to solve the non-convex trajectory optimization problems therein. Though providing a general framework for trajectory optimization in two-dimensional (2D) space, the studies in \cite{zeng2016throughput} and \cite{zeng2016energy} only focus on the setup with single UAV and single ground user. For UAV-enabled multi-user system,  a novel cyclical multiple access scheme is proposed in \cite{lyu2016cyclical}, where the UAV communicates with ground users when it flies sufficiently close to each of them in a periodic (cyclical) time-division manner.
An interesting throughput-access delay tradeoff is revealed and it has been shown that significant throughput gains can be achieved over the case of a static UAV for delay-tolerant applications. However,  only one single UAV with the constant flying speed is considered in \cite{lyu2016cyclical}, and the ground users are assumed to be uniformly located in a one-dimensional (1D) line, which simplifies the analysis but limits the applicability in practice.

In this paper, we study a general multi-UAV enabled wireless communication system, where multiple UAVs are employed to serve a group of users on the ground in a given 2D area.
Although a single UAV has demonstrated its advantages in performance enhancement for wireless networks  \cite{zeng2016throughput,zeng2016energy,wu2017joint, guangchi2016_UAV,JR:guangchi2016_UAV,wu2017ofdma,JR:wu2017_ofdm,yang2017energy},  it has limited capability in general and may not guarantee availability during the entire mission due to its practical size, weight and power (SWAP) constraints \cite{zeng2016wireless}.  This thus motivates the deployment of multiple or a swarm of UAVs which cooperatively serve the ground users to achieve more efficient communications.  For example, a group of UAVs may be deployed to keep track of the participants in a large-area event and to form a multi-hop communication network connecting to the ground audience.  More importantly, in a multi-UAV enabled network, users could be served in parallel with higher throughput and lower access delay, which could effectively alleviate the fundamental throughput-access delay tradeoff in single-UAV communications \cite{lyu2016cyclical}.

Without loss of generality, we consider that all UAVs share the same frequency band  for their communications with the ground users.
 By focusing on the downlink transmission from the UAVs to ground users, our goal is to maximize the minimum average rate among all users by jointly optimizing the user communication scheduling and association, and the UAV trajectory and transmit power control in a given finite period.
Such a joint optimization problem is practically appealing,  but has not been investigated in the literature to the authors'  best knowledge.
On one hand, by properly  designing the trajectories of different UAVs, not only short-distance LoS links can be proactively and dynamically established for those desired UAV-user pairs, but also the interfering channel distances between the undesired UAV-user pairs can be enlarged to
alleviate the co-channel interference. On the other hand, in the occasional scenarios when the UAVs have to get close with each other for serving nearby users,  their transmission power can be adjusted to reduce interference. While maximum transmission power is used for maximizing spectrum efficiency when the UAVs are far apart to serve users that are well separated. Therefore, the system performance can benefit from different design dimensions of the proposed joint optimization.
 However, such a joint  trajectory and adaptive communication design problem is non-trivial to solve. 
  This is because the user scheduling and association, UAV trajectory optimization, and transmit power control are closely coupled with each other in our considered problem, which makes it challenging to solve in general.

 To tackle the above challenges, we first relax the binary variables for user scheduling and association into continuous variables and solve the resulting problem with an efficient iterative algorithm  by leveraging the block coordinate descent method \cite{hong2016unified}.
 Specifically, the entire optimization variables are  partitioned into three blocks for the user scheduling and association, UAV trajectory, and transmit power control, respectively. Then, these three blocks of variables are alternately optimized in each iteration, i.e., one block is optimized at each time while keeping the other two blocks fixed.
  However, even with fixed user scheduling and association, the UAV trajectory optimization problem with fixed power control and the UAV power control problem with fixed trajectory are still difficult to solve due to their non-convexity. We thus apply the successive convex optimization technique to solve them approximately.
We also show that our proposed algorithm is guaranteed to converge. To speed up the algorithm convergence and achieve a superior performance, we propose an efficient and systematic trajectory initialization scheme based on the simple circular trajectory and the circle packing scheme. Numerical results show that significant throughput gains are achieved by our proposed joint design, as compared to conventional static UAV or other benchmark schemes with heuristic UAV trajectories. It is also shown that the throughput of the proposed mobile UAV system increases with the UAV trajectory design period, revealing the general throughput-access delay tradeoff \cite{wu2017joint,lyu2016cyclical} in multi-UAV enabled communications. In addition, compared to the single-UAV case, this  tradeoff is shown to be significantly improved by the use of multiple UAVs.

The rest of this paper is organized as follows. Section II introduces the system model and the problem formulation for a multi-UAV enabled wireless network.
In Section III, we propose an efficient iterative algorithm by applying the block coordinate descent and the successive convex optimization techniques.
Section VI presents the numerical results to demonstrate the performance of the proposed design. Finally, we conclude the paper  in Section VI.

\emph{Notations:} In this paper, scalars are denoted by italic letters, vectors and matrices are respectively denoted by bold-face lower-case and upper-case letters. $\mathbb{R}^{M\times 1}$ denotes the space of $M$-dimensional real-valued vector. For a vector $\mathbf{a}$, $\|\mathbf{a}\|$ represents its Euclidean norm and $\mathbf{a}^T$ denotes its transpose. For a time-dependent function $\mathbf{x}(t)$,  $\dot{\mathbf{x}}(t)$ denotes the derivative with respect to time $t$. For a set $\mathcal{K}$, $|\mathcal{K}|$ denotes its cardinality.
\section{System Model and Problem Formulation}

 \begin{figure}[!t]
\centering
\includegraphics[width=0.4\textwidth]{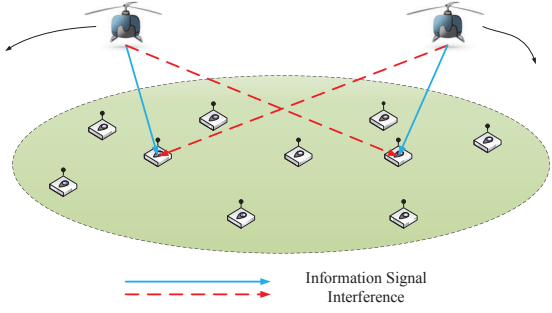}
\caption{A multi-UAV enabled wireless network. } \label{UAV}
\end{figure}
\subsection{System Model}
As shown in Fig. 1, we consider a wireless communication system where $M\geq 1 $ UAVs are employed to serve a group of $K>1$ ground users.  The user and UAV sets are denoted as $\mathcal{K}$ and $\mathcal{M}$, respectively, where $|\mathcal{K}|=K$ and $|\mathcal{M}|=M$. This practically corresponds to an information broadcast system enabled by UAVs.
  Assume that all the UAVs share the same frequency band for communication over consecutive periods each of duration  $T >0$ in second (s).  During any period,  each of the UAVs serves its associated ground users via a periodic/cyclical time-division multiple access (TDMA).
 Note that the choice of $T$ has a significant impact on the system performance. On one hand, thanks to the UAV mobility, a larger period  $T$ provides more time for each UAV to move closer to its served users  to achieve better communication channels, as well as to fly sufficiently away from the users served by other UAVs  for more effective interference mitigation,  thus achieving  a higher throughput.
{On the other hand, a larger $T$ in general implies a larger access delay for users since each user may need to wait for a longer time to be scheduled to communicate with a UAV between two periods.}
{Therefore, the period $T$ needs to be properly chosen in practice to strike a balance between the user throughput and access delay, i.e., there exists a fundamental throughput-access delay tradeoff \cite{lyu2016cyclical} in UAV-enabled communications.}

Without loss of generality, we consider a 3D Cartesian coordinate system where  the horizontal coordinate of each ground user $k$ is fixed at ${\mathbf{w}}_{k}=[x_k,y_k]^T \in \mathbb{R}^{2\times 1}$, $k\in \mathcal{K}$. All UAVs are assumed to fly at a fixed altitude $H$ above ground and the time-varying horizontal coordinate of UAV $m\in \mathcal{M}$ at time instant $t$ is denoted by $\mathbf{q}_m(t)=[x_m(t), y_m(t)]^T\in \mathbb{R}^{2\times 1}$, with $0\leq t\leq T$. The UAV trajectories need to satisfy the following constraint
\begin{align}
\q_m(0) &= \q_m(T), \forall\, m, \label{eq01}
\end{align}
which implies that each UAV needs to return to its initial location by the end of each period $T$ such that users can be served periodically in the next period.  { In practice, the trajectories of UAVs are also subject to the maximum speed constraints\footnote{Here, we do not consider the minimum speed constraints, which is practically valid for the rotary-wing UAVs with the capability of keeping stationary at fixed positions, i.e., a minimum zero-speed is feasible. However, for the fixed-wing UAVs that must move forward to remain aloft, additional minimum speed constraints, i.e., $||\dot{\q}_m(t)||\geq V_{\min}>0$, $0\leq t \leq T, \forall\, m$, need to be imposed  \cite{zeng2016energy}, which can be handled by the proposed algorithm with only a minor modification.} and collision avoidance constraints, i.e,
\begin{align}
||\dot{\q}_m(t)|| &\leq V_{\max},  0\leq t \leq T, \forall\, m, \label{eq02}\\
|| \q_m(t)-  \q_{j}(t)|| &\geq d_{\min},  0\leq t \leq T, \forall\, j\neq m, \label{eq0220}
\end{align}
where $V_{\max}$ in \eqref{eq02} denotes the maximum UAV speed in meter/second (m/s) and $d_{\min}$ denotes the minimum inter-UAV distance in m to ensure collision avoidance.}
For ease of exposition, the  period $T$ is  discretized into $N$ equal-time slots, indexed by $n=1,...,N$.  The elemental slot length $\delta_t = \frac{T}{N}$ is chosen to be sufficiently small such that a UAV's location is considered as approximately unchanged within each time slot even at the maximum speed $V_{\max}$. As a result, the trajectory of UAV $m$  can be approximated by the $N$ two-dimensional sequences $\mathbf{q}_m[n]=[x_m[n], y_m[n]]^T$, $n=1,\cdots, N$.  Furthermore, the trajectory constraints (\ref{eq01})--(\ref{eq0220}) can be equivalently written as
{
\begin{align}
\q_m[1] &= \q_m[N],\\
||\mathbf{q}_m[n+1]-\mathbf{q}_m[n]||^2 &\leq S_{\max}^2,  n=1,...,N-1, \\
|| \q_m[n]-  \q_{j}[n]||^2 &\geq d^2_{\min}, \forall\, n, m, j\neq m,  \label{eq022}
\end{align}}
\kern -1.6mm where $S_{\max} \triangleq V_{\max}\delta_t$ is the maximum horizontal distance that the UAV can travel in each time slot.
 In fact, any required accuracy of the adopted discrete-time approximation can be always satisfied by choosing a minimum $N$, as follows. To guarantee a certain accuracy, the ratio of $S_{\max}$ and $H$ can be restricted below a threshold, i.e., $\frac{S_{\max}}{H}\leq \varepsilon_{\max}$, where $\varepsilon_{\max}$ is the given threshold.  Then, the minimum number of time slots  required for achieving the accuracy with a given  $\varepsilon_{\max}$ can be obtained as
 \begin{align}
 N\geq \frac{V_{\max}T}{H \varepsilon_{\max}}.
 \end{align}
 However, further increasing  $N$ also increases our design complexity.  Therefore, the number of time slots $N$ can be properly chosen in practice to  balance between the accuracy and complexity.

The distance from UAV $m$  to user $k$ in time slot $n$  can be expressed as
 \begin{align}
 d_{k,m}[n] =  \sqrt{H^2 +||\q_m[n]-\w_k||^2}.
 \end{align}
 For simplicity, we assume that the communication links from the UAV to the ground users are dominated by the LoS links where the channel quality depends only on the UAV-user distance. Furthermore, the Doppler effect caused by the UAV mobility  is assumed to be well compensated at the  receivers.  Thus, the channel power gain from UAV $m$ to user $k$ during slot $n$ follows the free-space path loss model, which can be  expressed as
   \begin{align}
h_{k,m}[n]& = \rho_0d^{-2}_{k,m}[n] =\frac{\rho_0}{ H^2 +||\q_m[n]-\w_k||^2},
 \end{align}
  where $\rho_0$ denotes the channel power at the reference distance $d_0=1$ m.
 Define a binary  variable $\alpha_{k,m}[n]$, which indicates that user $k$ is served by UAV $m$ in time slot $n$ if $\alpha_{k,m}[n]=1$; otherwise, $\alpha_{k,m}[n]=0$.
 As such, $\alpha_{k,m}[n]$ specifies not only the user communication scheduling across the different time slots, but also the UAV-user association for each time slot.
  We assume that in each time slot, each UAV only serves at most one user and each user is only served by at most one UAV, which yields the following constraints
\begin{align}
&\sum_{k=1}^{K}\alpha_{k,m}[n]\leq 1, \forall\,m, n,        \label{eq77}   \\
& \sum_{m=1}^{M}\alpha_{k,m}[n]\leq 1, \forall\,k,  n,    \label{eq88} \\
 &\alpha_{k,m}[n]\in\{0, 1\}, \forall\, k, m, n.
\end{align}
The downlink transmit power of UAV $m$, $m\in \mathcal{M}$ in time slot $n$ is denoted by $p_m[n]$, which is subject to the constraint $0 \leq p_m[n]\leq P_{\max}$, with $P_{\max}$ denoting the peak UAV transmission power.  Thus, if user $k$ is served by UAV $m$ in time slot $n$, i.e., $\alpha_{k,m}[n]=1$,  the corresponding received signal-to-interference-plus-noise ratio (SINR) at user $k$ can be expressed as
  \begin{align}\label{eq2}
  \gamma_{k,m}[n] = \frac{p_m[n]h_{k,m}[n]}{\sum_{j=1, j\neq m}^{M}p_j[n]h_{k,j}[n]+\sigma^2},
  \end{align}
where $\sigma^2$ is the power of the additive white Gaussian noise (AWGN) at the receiver.
 The term $\sum_{j=1, j\neq m}^{M}p_j[n]h_{k,j}[n]$ in the denominator of (\ref{eq2}) represents the co-channel interference caused by the transmissions of all other UAVs in time slot $n$.
Thus, the achievable rate of user $k$ in time slot $n$, denoted by $R_{k}[n]$ in bits/second/Hertz (bps/Hz), can be expressed as
 \begin{align}
R_{k}[n] = \sum_{m=1}^{M}\alpha_{k,m}[n]  \log_2\left( 1 +  \gamma_{k,m}[n] \right).
 \end{align}
Thus, the achievable average rate of user $k$ over $N$ time slots is given by
$R_k= \frac{1}{N}\sum_{n=1}^{N}R_{i}[n].$


\subsection{Problem Formulation}
Let $\mathbf{A}=\{\alpha_{k,m}[n], \forall\, k,m,n\}$, $\mathbf{Q}=\{\mathbf{q}_{m}[n], \forall\,m,n\}$, and $\mathbf{P}=\{{p}_{m}[n], \forall\,m,n\}$.
By assuming that the locations of the ground users are known, our goal is to maximize the minimum average rate among all users by jointly optimizing the user scheduling and association (i.e., $\A$), UAV trajectory (i.e., $\Q$), and transmit power  (i.e., $\pow$) over all time slots. Define $\eta(\A,\Q, \pow)= \min \limits_{k\in \mathcal{K}}~ R_k$ as a function of $\A$, $\Q$, and $\pow$. The optimization problem is formulated as
\begin{subequations}  \label{probm6}
 \begin{align}
 &\max  \limits_{\eta,\mathbf{A},\mathbf{Q}, \mathbf{P}} ~ \eta  \label{probm06}  \\
&\text{s.t.} ~\frac{1}{N} \sum_{n=1}^{N}\sum_{m=1}^{M} \alpha_{k,m}[n]  \log_2\left( 1 +  \gamma_{k,m}[n] \right) \geq \eta, \forall\, k, \label{eq012} \\
 &~~~~ \sum_{k=1}^{K}\alpha_{k,m}[n]\leq 1, \forall\,m, n,        \label{eq7}   \\
&~~~~  \sum_{m=1}^{M}\alpha_{k,m}[n]\leq 1, \forall\,k,  n,    \label{eq8}  \\
&~ ~~~  \alpha_{k,m}[n]\in\{0, 1\}, \forall\, k,m,  n,     \label{eq9}  \\
&~ ~~~   ||\mathbf{q}_m[n+1]-\mathbf{q}_m[n]||^2\leq S_{\max}^2,  n=1,...,N-1,    \label{eq10} \\
&~~~~   \mathbf{q}_m[1]=\mathbf{q}_m[N],   \forall\, m,      \label{eq11} \\
&~ ~~~  {|| \q_m[n]-  \q_{j}[n]||^2 \geq d^2_{\min},   \forall\, n, m, j\neq m, }\label{eq11216}\\
&~ ~~~ 0\leq p_m[n]\leq P_{\max}, \forall\,m, n.   \label{eq12} 
 \end{align}
 \end{subequations}

 Problem  (\ref{probm6}) is challenging to solve due to the following two main reasons. First, the optimization variables $\mathbf{A}$ for user scheduling and association are binary and thus (\ref{eq7})-(\ref{eq9}) involve integer constraints. Second, even with fixed user scheduling and association,  (\ref{eq012}) and \eqref{eq11216} are still non-convex constraints with respect to UAV trajectory variables $\Q$ and/or transmit power variables $\pow$.
 Therefore, problem  (\ref{probm6}) is a mixed-integer non-convex problem, which is difficult to be optimally solved in general.
\section{Proposed Algorithm}

To make problem (\ref{probm6}) more tractable, we first relax the binary variables in (\ref{eq9}) into continuous variables, which yields the following problem
  \begin{subequations}\label{probm66}
 \begin{align}
&\max  \limits_{\eta,\mathbf{A},\mathbf{Q},\pow} ~~~ ~\eta  \\ 
&~~\text{s.t.}  ~~ 0\leq\alpha_{k,m}[n]\leq 1, \forall\, k,m,n,   \label{eq13d} \\
&~~~~~~ ~ \text{(\ref{eq012}),  (\ref{eq7}),  (\ref{eq8}), (\ref{eq10}), (\ref{eq11}),  \eqref{eq11216}, (\ref{eq12})}.   \label{eq13d}
 \end{align}
 \end{subequations}
 Such a relaxation in general implies that the objective value of problem (\ref{probm66}) serves as an upper bound for that of problem  (\ref{probm6}).
 Although relaxed, problem (\ref{probm66}) is still a non-convex optimization problem due to the non-convex constraint (\ref{eq012}). In general, there is no standard method for solving such non-convex optimization problems efficiently. 
  In the following, we  propose an efficient iterative algorithm for the relaxed problem  (\ref{probm66}) by applying the block coordinate descent  \cite{hong2016unified} and successive convex optimization techniques. Specifically, for given UAV trajectory $\mathbf{Q}$ and transmit power $\pow$, we optimize the user scheduling and association $\A$ by solving a linear programming (LP). For any given user scheduling and association $\A$ and transmit power $\pow$ (UAV trajectory $\mathbf{Q}$), the UAV trajectory $\Q$ (transmit power $\pow$) is optimized based on the successive convex optimization technique \cite{zeng2016throughput,zeng2016energy}.
 Then, we present the overall algorithm and analytically show its convergence. Furthermore, we propose a low-complexity  initialization scheme for the UAV trajectory design. Finally, we show how to reconstruct a binary solution to the original problem (\ref{probm6}) based on the obtained solution to problem  (\ref{probm66}).
 \subsection{User Scheduling and Association Optimization}
 For any given UAV trajectory and transmit power  $\{\Q, \mathbf{P}\}$, the user scheduling and association of problem (\ref{probm66}) can be optimized by solving the following problem
 \begin{subequations}\label{probm25}
  \begin{align}
&\mathop {\text{max} }\limits_{\eta,\A}~~ \eta   \label{probm250}\\
&~\text{s.t.}~   \frac{1}{N} \sum_{n=1}^{N} \sum_{m=1}^{M} \alpha_{k,m}[n]  \log_2\left( 1 + \gamma_{k,m}[n]  \right)  \geq \eta, \forall\, k, \label{eq26}\\
&~~~~~\sum_{k=1}^{K}\alpha_{k,m}[n]\leq 1, \forall\,m, n,        \label{eq70}   \\
&~~~~~  \sum_{m=1}^{M}\alpha_{k,m}[n]\leq 1, \forall\,k,  n,    \label{eq80}  \\
&~~ ~~~   0\leq\alpha_{k,m}[n]\leq 1, \forall\, k,m,  n.     \label{eq90}
 \end{align}
  \end{subequations}
Since problem (\ref{probm25}) is a standard LP, it can be solved efficiently by existing  optimization tools such as CVX \cite{cvx}. Furthermore,  it is easy to see that the constraints (\ref{eq70}) and (\ref{eq80}) are met with equalities when the optimal solution $\A$ is attained for given $\{\Q, \mathbf{P}\}$.
%
  \subsection{UAV Trajectory Optimization}
%
 For any given user scheduling and association as well as UAV transmit power \{$\A$, $\pow$\},  the UAV trajectory of problem (\ref{probm66}) can be optimized by solving the following problem
  \begin{subequations} \label{probm55}
    \begin{align}
 &\mathop {\text{max} }\limits_{\eta,\Q}~ \eta \label{probm550}  \\ %
&~\text{s.t.}  \frac{1}{N}\sum_{n=1}^{N}\sum_{m=1}^{M}\alpha_{k,m}[n] \log_2\left( 1 +  \gamma_{k,m}[n] \right)  \geq \eta, \forall\, k,\label{eq152} \\
&~~~~   ||\mathbf{q}_m[n+1]-\mathbf{q}_m[n]||^2\leq S_{\max}^2,  n=1,...,N-1,    \label{eq100} \\
&~~~~  \mathbf{q}_m[1]=\mathbf{q}_m[N],   \forall\, m,      \label{eq110} \\
&~~~~ || \q_m[n]-  \q_j[n]||^2 \geq d^2_{\min},   \forall\, n, m, j\neq m. \label{eq112}
 \end{align}
 \end{subequations}
Note that problem (\ref{probm55}) is neither a concave or quasi-concave maximization problem due to the non-convex constraints in (\ref{eq152}) and \eqref{eq112}. In general, there is no efficient method to obtain the optimal solution. 
In the following, we adopt the successive convex optimization technique for the trajectory optimization. To this end, ${R}_{k,m}[n]$, in constraints (\ref{eq152}) can be written as
\begin{align}\label{eq291}
{R}_{k,m}[n]&=   \log_2\left( 1 +\frac{\frac{p_m[n]\rho_0}{H^2+||\mathbf{q}_m[n]-\mathbf{w}_k||^2}}{\sum_{j=1, j\neq m}^{M}\frac{p_j[n]\rho_0}{H^2+||\mathbf{q}_j[n]-\mathbf{w}_k||^2}+\sigma^2} \right)   \nonumber\\
&\kern -12mm =  \hat{R}_{k,m}[n]- \log_2\left(\sum_{j=1, j\neq m}^{M}\frac{p_j[n]\rho_0}{H^2+||\mathbf{q}_j[n]-\mathbf{w}_k||^2}+\sigma^2\right),
\end{align}
where
\begin{align}
  \hat{R}_{k,m}[n] &=   \log_2\left(\sum_{j=1}^{M} \frac{p_j[n]\rho_0}{H^2+||\mathbf{q}_j[n]-\mathbf{w}_k||^2}+ \sigma^2 \right). \label{eq29}
  \end{align}
  With (\ref{eq291}) and (\ref{eq29}), constraints (\ref{eq152}) are transformed into
  \begin{align}
 &\frac{1}{N}\sum_{n=1}^{N}\sum_{m=1}^{M}\alpha_{k,m}[n]\Bigg(  \hat{R}_{k,m}[n]-  \nonumber \\
& \log_2\bigg(\sum_{j=1, j\neq m}^{M}\frac{p_j[n]\rho_0}{H^2+||\mathbf{q}_j[n]-\mathbf{w}_k||^2}+\sigma^2\bigg)\Bigg)  \geq \eta, \forall\, k. \label{eq22b}
  \end{align}
 Note that constraints in (\ref{eq22b}) are still non-convex. By introducing slack variables $\mathbf{S}=\{S_{k,j}[n]=||\mathbf{q}_j[n]-\mathbf{w}_k||^2,\forall\, j\neq m, j\in\mathcal{M}, k, n\} $, problem (\ref{probm55}) can be reformulated as
 \begin{subequations}\label{probm34}
   \begin{align} 
&\kern -3mm \mathop {\text{max} }\limits_{\eta,\mathbf{Q}, \mathbf{S}}~ \eta  \\
&\kern -3mm \text{s.t.} ~~ \frac{1}{N} \sum_{n=1}^{N}\sum_{m=1}^{M}\alpha_{k,m}[n] \Bigg(  \hat{R}_{k,m}[n] \nonumber\\
&- \log_2\bigg(\sum_{j=1, j\neq m}^{M} \frac{p_j[n]\rho_0}{H^2+S_{k,j}[n]}+\sigma^2\bigg)\Bigg) \geq \eta, \forall\, k, \label{eq35} \\
&S_{k,j}[n] \leq ||\mathbf{q}_j[n]-\mathbf{w}_k||^2,  \forall\, k, j \neq m, n, \label{eq36}\\
&||\mathbf{q}_m[n+1]-\mathbf{q}_m[n]||^2\leq S_{\max}^2,  n=1,...,N-1,    \label{eq100} \\
&\mathbf{q}_m[1]=\mathbf{q}_m[N],   \forall\, m,      \label{eq110} \\
&|| \q_m[n]-  \q_j[n]||^2 \geq d^2_{\min},   \forall\, n, m, j\neq m. \label{eq11220}
 \end{align}
  \end{subequations}
It can be verified that without loss of optimality to problem (\ref{probm34}), all constraints in (\ref{eq36}) can be met with equality, since otherwise we can always increase $S_{k,j}[n]$ without decreasing the objective value. Note that in (\ref{eq35}), $\hat{R}_{k,m}[n]$ is neither convex nor concave with respect to $\mathbf{q}_j[n]$. While in (\ref{eq36}), even though $||\mathbf{q}_j[n]-\mathbf{w}_k||^2$ is convex with respect to $\mathbf{q}_j[n]$, the resulting set is not  a convex set since the superlevel set of a convex quadratic function is not convex in general. Thus,  problem (\ref{probm34}) is still a non-convex optimization problem due to the non-convex feasible set.


To tackle the non-convexity of  (\ref{eq35}), (\ref{eq36}), and \eqref{eq11220}, the successive convex optimization technique can be applied where in each iteration, the original function is approximated by a more tractable function at a given local point. Specifically, define $\Q^r =\{\mathbf{q}_m^r[n], \forall\,m, n\}$ as the given trajectory of UAVs in the $r$-th iteration\footnote{{In Section III-D, we show that $\Q^r$ is in fact the solution obtained from the $(r-1)$th iteration.  } }.  The key observation is that in (\ref{eq29}),  although  $\hat{R}_{k,m}[n]$  is not concave with respect to $\mathbf{q}_j[n]$, it  is convex with respect to $||\mathbf{q}_j[n]-\mathbf{w}_k||^2$.  Recall that any convex function is globally lower-bounded by its first-order Taylor expansion at any point  \cite{Boyd}.  Therefore,  with given local point $\Q^r$ in the $r$-th iteration, we obtain the following  lower bound for  $\hat{R}_{k,m}[n]$ as in  \cite{zeng2016throughput,zeng2016energy}, i.e.,


{ \begin{align}
 \hat{R}_{k,m}[n] &=   \log_2\left(\sum_{j=1}^{M} \frac{p_j[n]\rho_0}{H^2+||\mathbf{q}_j[n]-\mathbf{w}_k||^2}+ \sigma^2 \right)
 \nonumber\\
&\kern -1cm \geq \sum_{j=1}^{M}-A^r_{k,j}[n]\left(||\mathbf{q}_j[n]-\mathbf{w}_k||^2 -||\mathbf{q}^r_j[n]-\mathbf{w}_k||^2 \right) \nonumber \\
& \kern -0.5cm+  B^r_{k,j}[n]  \triangleq \hat{R}^{\rm lb}_{k,m}[n],  \label{eq37}
\end{align} }
where $A^r_{k,j}[n]$ and $B^r_{k,j}[n]$ are constants that are given by
\begin{align}
A^r_{k,j}[n]&= \frac{ \frac{p_j[n]\rho_0}{(H^2+||\mathbf{q}_j^r[n]-\mathbf{w}_k||^2)^2}\log_2(e)}{\sum_{l=1}^{M} \frac{p_l[n]\rho_0}{H^2+||\mathbf{q}_l^r[n]-\mathbf{w}_k||^2} + \sigma^2  }, \forall\, k,j, n, \\
B^r_{k,j}[n]&= \log_2\left( \sum_{l=1}^{M}\frac{p_l[n]\rho_0}{H^2+||\mathbf{q}_l^r[n]-\mathbf{w}_k||^2}+\sigma^2 \right)  , \forall\, k,j, n.
\end{align}
In constraints (\ref{eq36}), since $||\mathbf{q}_j[n]-\mathbf{w}_k||^2$ is a convex function with respect to $\mathbf{q}_j[n]$,
we have the following inequality by applying the first-order Taylor expansion at the given point $\mathbf{q}_j^r[n]$,
\begin{align}
 & ||\mathbf{q}_j[n]-\mathbf{w}_k||^2\geq    |\mathbf{q}_j^r[n]-\mathbf{w}_k||^2  \nonumber\\
 &~~~~~+  2(\mathbf{q}_j^r[n] - \mathbf{w}_k)^T( \mathbf{q}_j[n] - \mathbf{q}_j^r[n] ), \forall\, k,j\neq m,n. \label{eq40}
\end{align}
{Similarly,  by applying the first-order Taylor expansion at the given point $\mathbf{q}_m^r[n]$ and $\mathbf{q}_j^r[n]$ to $|| \q_m[n]-  \q_j[n]||^2$, we obtain
\begin{align}
& ||\mathbf{q}_m[n]-\mathbf{q}_j[n]||^2 \geq    -||\mathbf{q}_m^r[n]-\mathbf{q}_j^r[n]||^2 \nonumber\\
 &~~~+  2(\mathbf{q}_m^r[n] - \mathbf{q}_j^r[n])^T( \mathbf{q}_m[n]-\mathbf{q}_j[n] ), \forall\, j\neq m,n. \label{eq400}
\end{align}}

 With any given local point $\Q^r$ as well as  the lower bounds  in  (\ref{eq37}) and (\ref{eq40}),  problem (\ref{probm34}) is approximated as the following problem 
 \begin{subequations} \label{probm41}
   \begin{align}
  &\kern -2mm \mathop {\text{max} }\limits_{\eta^{r}_{\rm trj},\mathbf{Q},\mathbf{S}}~~\eta^{r}_{\rm trj} \\
&\kern -2mm\text{s.t.} ~  \frac{1}{N}\sum_{n=1}^{N}\sum_{m=1}^{M}\alpha_{k,m}[n] \Bigg(  \hat{R}^{\rm lb}_{k,m}[n] \nonumber\\
&~~~- \log_2\bigg(\sum_{j=1, j\neq m}^{M} \frac{p_j[n]\rho_0}{H^2+S_{k,j}[n]}+\sigma^2\bigg)\Bigg) \geq \eta^{ r}_{\rm trj}, \forall\, k,   \label{eq41}\\
&\kern -2mm  S_{k,j}[n] \leq   ||\mathbf{q}_j^r[n]-\mathbf{w}_k||^2 \nonumber\\
&~~~+  2(\mathbf{q}_j^r[n] - \mathbf{w}_k)^T( \mathbf{q}_j[n] - \mathbf{q}_j^r[n] ),  \forall\, k, j \neq m, n,  \label{eq42} \\
&\kern -2mm  ||\mathbf{q}_m[n+1]-\mathbf{q}_m[n]||^2\leq S_{\max}^2,  n=1,...,N-1,    \label{eq100} \\
&\kern -2mm \mathbf{q}_m[1]=\mathbf{q}_m[N],   \forall\, m,     \label{eq110} \\
&\kern -2mm  d^2_{\min} \leq    -||\mathbf{q}_m^r[n]-\mathbf{q}_j^r[n]||^2 \nonumber\\
&\kern -2mm +  2(\mathbf{q}_m^r[n] - \mathbf{q}_j^r[n])^T( \mathbf{q}_m[n]-\mathbf{q}_j[n] ), \forall\, n, m, j\neq m. \label{eq110new}
 \end{align}
  \end{subequations}
  Since the left-hand-side (LHS) of the constraint (\ref{eq41}) is jointly concave with respect to $\mathbf{q}_j^r[n]$ and $S_{k,j}[n]$,  it is convex now.  {Furthermore, (\ref{eq100}) is a convex quadratic constraint and (\ref{eq42}),  (\ref{eq110}), and \eqref{eq110new} are all linear constraints.} Therefore, problem (\ref{probm41}) is  a convex optimization problem that can be efficiently solved by standard convex optimization solvers such as  CVX  \cite{Boyd}. It is worth noting that the lower bounds adopted in (\ref{eq41}) and (\ref{eq42}) suggest that any feasible solution of problem (\ref{probm41}) is also feasible for problem (\ref{probm34}), but the reverse does not hold in general. As a result, the optimal objective value obtained from the approximate problem (\ref{probm41}) in general serves as a lower bound of that of problem (\ref{probm34}).
\subsection{UAV Transmit Power Control}
 For any given user scheduling and association as well as UAV trajectory \{$\A$, $\Q$\},  the UAV transmit power of problem (\ref{probm66}) can be optimized by solving the following problem
 \begin{subequations} \label{probm44}
  \begin{align}
 &~\mathop {\text{max} }\limits_{\eta,\mathbf{P}}~~ \eta  \\
&~\text{s.t.} ~  \frac{1}{N}\sum_{n=1}^{N}\sum_{m=1}^{M}\alpha_{k,m}[n]  \log_2\left( 1 +  \gamma_{k,m}[n] \right)   \geq \eta, \forall\, k, \label{eq45}\\
&~~~~~0\leq p_m[n]\leq P_{\max}, \forall\,m, n. \label{eq455}
 \end{align}
 \end{subequations}
 Problem (\ref{probm44}) is a non-convex optimization problem due to the non-convex constraint (\ref{eq45}) and in fact NP-hard for general $N$.
 Note that the LHS of (\ref{eq45}), i.e.,  ${R}_{k,m}[n]$, can be written as a difference of two concave functions with respect to the power control variables, i.e.,
\begin{align}
{R}_{k,m}[n]&=    \log_2\left( 1 +\frac{p_m[n]h_{k,m}[n]}{\sum_{j=1, j\neq m}^{M}p_j[n]h_{k,j}[n]+\sigma^2} \right)   \nonumber\\
&=    \log_2\left(\sum_{j=1}^{M} p_j[n]h_{k,j}[n]+ \sigma^2 \right) -   \check{R}_{k,m}[n], \label{eq370}
\end{align}
where
\begin{align}
  \check{R}_{k,m}[n] &= \log_2\left(\sum_{j=1, j\neq m}^{M}p_j[n]h_{k,j}[n]+\sigma^2\right).
  \end{align}
To handle the non-convex contraint of  (\ref{eq45}),  we apply the successive convex optimization technique to approximate $\check{R}_{k,m}[n]$ with a convex function in each iteration. Specifically, define $\pow^r =\{\mathbf{p}_m^r[n], \forall\,m, n\}$ as the given transmit power of UAV $m$ in the $r$-th iteration.   Recall that any concave function is globally upper-bounded by its first-order Taylor expansion at any point \cite{Boyd}.  Thus, we have the following convex upper bound at the given local point $p^r_j[n]$
\begin{align}
 \check{R}_{k,m}[n] &= \log_2\left(\sum_{j=1, j\neq m}^{M}p_j[n]h_{k,j}[n]+\sigma^2\right)
 \nonumber\\
& \leq \sum_{j=1, j\neq m}^{M} D_{k,j}[n]\left(p_j[n]-p_j^r[n] \right) \nonumber\\
&~~~~~~ + \log_2\left( \sum_{j=1, j\neq m}^{M}p_j^r[n]h_{k,j}[n]+\sigma^2 \right)  \nonumber \\
&\triangleq \check{R}^{\rm ub}_{k,m}[n],  \label{eq48}
\end{align}
where 
\begin{align}
D_{k,j}[n]=  \frac{h_{k,j}[n]\log_2(e)}{ \sum_{l=1, l\neq m}^{M}p_j^r[n]h_{k,l}[n]+\sigma^2},  \forall\, k,j, n. 
\end{align}

With any given local point  $\pow^r$ and  the upper bound $\check{R}^{\rm ub}_{k,m}[n]$  in (\ref{eq48}), problem (\ref{probm44})  is approximated as the following problem
\begin{subequations}\label{probm50}
  \begin{align}
 &~\mathop {\text{max} }\limits_{\eta^r_{\rm pow},\mathbf{P}}~ \eta^r_{\rm pow}   \\
&\text{s.t.} ~   \frac{1}{N}\sum_{n=1}^{N}\sum_{m=1}^{M}\alpha_{k,m}[n]\Bigg(   \log_2\bigg(\sum_{j=1}^{M} p_j[n]h_{k,j}[n]+ \sigma^2 \bigg) \nonumber\\
& ~~~~~~~~~~ -\check{R}^{\rm ub}_{k,m}[n] \Bigg) \geq \eta^r_{\rm pow}, \forall\, k, \label{eq50}\\
&~~0\leq p_m[n]\leq P_{\max}, \forall\,m, n. \label{eq455}
 \end{align}
 \end{subequations}
Problem (\ref{probm50}) is a convex optimization problem, which  can be efficiently solved  by standard convex optimization solvers such as CVX \cite{Boyd}. It is also worth noting that the upper bound adopted in (\ref{eq50}) suggests that the feasible set of problem (\ref{probm50}) is always a subset of that of problem (\ref{probm44}).  Therefore, the optimal objective value obtained from problem (\ref{probm50}) in general serves as a lower bound of that of problem (\ref{probm44}).

\subsection{Overall Algorithm and Convergence}
%
Based on the results presented in the previous three subsections, we propose an overall  iterative algorithm for problem (\ref{probm66}) by applying the block coordinate descent method  \cite{bertsekas1999nonlinear}, also known as the alternating optimization method. Specifically, the entire optimization variables in original problem (\ref{probm66}) are partitioned into three blocks, i.e., $\{ \mathbf{A}, \Q, \mathbf{P}\}$. Then, the user scheduling and association $\A$, UAV trajectory $\Q$, and transmit power $\pow$ are alternately optimized, by  solving  problem (\ref{probm25}), (\ref{probm41}), and (\ref{probm50})  correspondingly, while keeping the other two blocks of variables fixed. {Furthermore, the obtained solution in each iteration is used as the input of the next iteration.}
The details of this algorithm are summarized in Algorithm 1.
It is worth pointing out that in the classical block coordinate descent method, the sub-problem for updating each block of variables is required to be solved exactly with optimality  in each iteration  in order to guarantee the convergence \cite{bertsekas1999nonlinear}. However, in our case, for the trajectory optimization problem (\ref{probm55}) and transmit power optimization problem  (\ref{probm44}), we only solve their approximate problems (\ref{probm41}) and (\ref{probm50}) optimally.  Thus, the convergence analysis for the classical  coordinate descent method cannot be directly applied and the convergence of Algorithm 1 needs to be proved, as shown next.


 \begin{algorithm}[t]
\caption{ Block coordinate descent algorithm for problem (\ref{probm66}).}\label{Algo:succ}
\begin{algorithmic}[1]
\STATE Initialize $\Q^0$ and $\mathbf{P}^0$. Let $r=0$.
\REPEAT
\STATE Solve problem (\ref{probm25}) for  given $\{\Q^r, \mathbf{P}^r\}$, and denote the optimal solution as $\{\mathbf{A}^{r+1}\}$.
\STATE {Solve problem (\ref{probm41}) for given $\{ \mathbf{A}^{r+1},\Q^r, \mathbf{P}^r\}$, and denote the optimal solution as $\{\Q^{r+1}\}$.}
\STATE Solve problem (\ref{probm50}) for given $\{\mathbf{A}^{r+1}, \Q^{r+1}, \mathbf{P}^r\}$, and denote the optimal solution as $\{\mathbf{P}^{r+1}\}$.
\STATE Update $r=r+1$.
\UNTIL{  The fractional increase of the objective value  is below a threshold $\epsilon>0$.}
\end{algorithmic}
\end{algorithm}

Define $\eta^{{\rm lb},r}_{\rm trj}(\A, \Q, \pow)=\eta^r_{\rm trj} $ and  $\eta^{{\rm lb},r}_{\rm pow}(\A, \Q, \pow)=\eta^r_{\rm pow} $ where $\eta^r_{\rm trj} $ and $\eta^r_{\rm pow}$ are respectively the objective values of problem (\ref{probm41}) and  (\ref{probm50}) based on $\A$, $\Q$, and $\pow$. First, in step 3 of Algorithm 1,  since the optimal solution of  (\ref{probm25}) is obtained for given $\Q^r$ and $\pow^r$, we have
\begin{align}\label{increase1}
\eta(\A^r, \Q^r, \pow^r) \leq \eta(\A^{r+1}, \Q^r, \pow^r),
\end{align}
where $\eta(\A, \Q, \pow)$ is defined prior to problem (\ref{probm6}).  Second, for given $\A^{r+1}$, $\Q^r$, and  $\pow^r$ in step 4 of Algorithm 1, it follows that
{\begin{align}\label{increase2}
\eta(\A^{r+1}, \Q^{r}, \pow^{r}) &\overset{(a)} = \eta^{{\rm lb}, r}_{\rm trj}(\A^{r+1}, \Q^{r}, \pow^{r}) \nonumber\\
&\overset{(b)} \leq \eta^{{\rm lb}, r}_{\rm trj}(\A^{r+1}, \Q^{r+1}, \pow^{r})\nonumber \\
&\overset{(c)} \leq \eta(\A^{r+1}, \Q^{r+1}, \pow^{r}),
\end{align}}
where $(a)$ holds since the first-order Taylor expansions in (\ref{eq37}) and (\ref{eq40}) are tight at the given local points, respectively,  which means that problem (\ref{probm41})  at  $\Q^r$ has the same objective  value as that of problem  (\ref{probm55});   $(b)$ holds since in step 4 of Algorithm 1 with the given $\A^{r+1}$ and $\pow^{r}$, problem (\ref{probm41}) is solved optimally with solution $\Q^{r+1}$;  $(c)$ holds since the objective  value of problem (\ref{probm41}) is the lower bound of that of its original problem (\ref{probm55}) at $\Q^{r+1}$.
%
The inequality in (\ref{increase2}) suggests that although only an approximate optimization problem (\ref{probm41}) is solved for obtaining the UAV trajectory, the objective  value of problem (\ref{probm55}) is still non-decreasing after each iteration.
{Third, for given $\A^{r+1}$, $\Q^{r+1}$, and $\pow^r$ in step 5 of Algorithm 1, it follows that
\begin{align}\label{increase3}
\eta(\A^{r+1}, \Q^{r+1}, \pow^{r}) &= \eta^{{\rm lb}, r}_{\rm pow}(\A^{r+1}, \Q^{r+1}, \pow^{r}) \nonumber\\
&\leq \eta^{{\rm lb}, r}_{\rm pow}(\A^{r+1},\Q^{r+1}, \pow^{r+1})\nonumber \\
& \leq \eta(\A^{r+1}, \Q^{r+1}, \pow^{r+1}),
\end{align}
which can be similarly shown as in \eqref{increase2}.
Based on (\ref{increase1})--(\ref{increase3}), we obtain
\begin{align}
\eta(\A^{r}, \Q^{r}, \pow^{r})\leq \eta(\A^{r+1}, \Q^{r+1}, \pow^{r+1}),
\end{align}
which indicates that the objective value of problem (\ref{probm66}) is non-decreasing after each iteration of Algorithm 1. } Since the objective value of problem  (\ref{probm66}) is upper bounded by a finite value, the proposed Algorithm 1 is guaranteed to converge. {Simulation results in Section IV show that the proposed block coordinate descent method converges quickly for our considered setup. Furthermore, since only convex optimization problems need to be solved  in each iteration of Algorithm 1, which are of polynomial complexity, Algorithm 1 can be practically implemented with fast convergence for wireless networks of a moderate number of users. }

Note that in Algorithm 1, the UAV trajectory has to be initialized. It is known that for such iterative algorithms, the converged solution and the ultimate system performance in general depend on the initialization schemes.  Thus, we further propose an efficient trajectory initialization scheme, which is elaborated in the next subsection.  

    \begin{figure}[!t]
\centering
\includegraphics[width=0.4\textwidth]{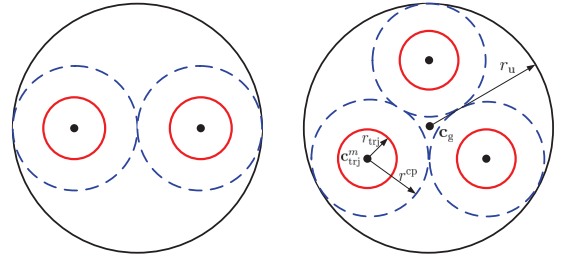}
\caption{An example of  UAV trajectories initialization based on circle packing  for $M=2$ (left) and $M=3$ (right). The black dots and the dashed blue circles are the results obtained from the circle packing scheme. The solid red circles are the initial circular trajectories of UAVs.}
\end{figure}
 \subsection{Trajectory Initialization Scheme}
In this subsection, we propose a low-complexity and systematic  initialization scheme for the trajectory design in Algorithm 1 based on the simple circular trajectory and the circle packing scheme. 
Specifically, the initial  trajectory of each UAV is set to be a circular trajectory with the UAV speed taking a constant value $V$, with $0< V\leq V_{\max}$. Furthermore, the radius of the initial trajectory circles are assumed to be the same for all UAVs. The center and radius of the circular trajectories are denoted by $\mathbf{c}^m_{\rm trj}=[x^m_{\rm trj}, y^m_{\rm trj}]^T$ and $r_{\rm trj}$, respectively. Thus, for any given period $T$, we have $2\pi r_{\rm trj}=VT$.
Intuitively, circles that correspond to the initial trajectories of different UAVs should be sufficiently separated to minimize the co-channel interference, and at the same time, all circles together should cover the entire area as much as possible so as to better balance the users' rates.
Therefore, the initial circular trajectories are obtained based on  circle packing. To this end, we first determine the geometric center of users as $\mathbf{c}_{\rm g}=\frac{\sum_{k=1}^{K}\mathbf{w}_k}{K}$. The radius of the minimum circle with $\mathbf{c}_{\rm g}$ as the circle center which can cover all users is denoted by $r_{\rm u}$, which is equal to the maximum distance between  $\mathbf{c}_{\rm g}$ and all the users, i.e.,
$r_{\rm u} = \max\limits_{k\in \mathcal K } ||\mathbf{w}_k-\mathbf{c}_{\rm g}||.$
Given the number of UAVs $M$ and   $r_{\rm u}$, we exploit the circle packing (CP) scheme \cite{circle_packing}, also known as point packing, to obtain the center of each of the $M$ circles $\mathbf{c}^m_{\rm trj}$ as well as the corresponding radius $r^{\rm cp}$. To balance the number of users inside and outside the circular trajectory, $\frac{r^{\rm cp}}{2}$ is a reasonable choice for the trajectory circle radius. However, due to the maximum UAV speed constraint, the resulting radius $\frac{r^{\rm cp}}{2}$ may not be always achievable given the finite time $T$ if $\pi r^{\rm cp}>V_{\max}T$. In this case, the maximum allowed radius is computed as
\begin{align}
r_{\max} = \frac{V_{\max}T}{2\pi}.
\end{align}
As such, the radius of the initial circular trajectory is set as $r_{\rm trj}=\min (r_{\max},\frac{r_{\rm cp} }{2} )$.  Let $\theta_n \triangleq  2\pi\frac{(n-1)}{N-1}$, $\forall\, n$, and  $\mathbf{Q}^0=\{\mathbf{q}^0_{m}[n], \forall\,m,n\}$. Based on $\mathbf{c}^m_{\rm trj}$ and $r_{\rm trj}$, the initial trajectory of UAV $m$ in time slot $n$ is then obtained as
\begin{align}\label{inital_trj}
\q^0_m[n] = \left[x^m_{\rm trj} + r_{\rm trj}\cos\theta_n,  y^m_{\rm trj} + r_{\rm trj}\sin\theta_n\right]^{T}, n=1,...,N.
\end{align}
Note that for $M\geq 2$, if the inter-UAV distance is larger than or equal to $d_{\min}$, then the trajectory obtained in \eqref{inital_trj} is feasible for original problem \eqref{probm6}. Otherwise, a feasible initial trajectory can be always obtained by scaling $r_{\rm u}$ such that $r_{\rm cp}$ is larger than or equal to $d_{\min}$.

  \subsection{Reconstruct the Binary User Scheduling and Association Solution}
  Note that Algorithm 1 is to solve the relaxed problem (\ref{probm66}) where the binary user scheduling and association variables in the original problem  (\ref{probm6}) are relaxed to continuous variables between 0 and 1.
Thus, in the solution obtained by Algorithm 1, if the user scheduling and association variables $\alpha_{k,m}[n]$ are all binary, then the relaxation is tight and the obtained solution is also a feasible solution of problem (\ref{probm6}). Otherwise, the binary user scheduling and association solution needs to be reconstructed based on the solution obtained for (\ref{probm66}).  To this end, we further divide each time slot into $\tau$ sub-slots so that the new total number of sub-slots is $N' = \tau N$, $\tau\geq 1$. Then, the number of sub-slots assigned to user $k$ by UAV $m$ in time slot $n$ is  $N_{k,m}[n]=\lfloor\tau \alpha_{k,m}[n]\rceil$, where $\lfloor x\rceil$ denotes the nearest integer of $x$.
It is not difficult to see that as $\tau$ increases, $N_{k,m}[n]$ approaches an integer which allows a binary solution.
For example, consider a single-UAV enabled two-user system with $\alpha_1[\ell]=0.69$ and $\alpha_2[\ell]=0.31$ in time slot $\ell$, where the UAV index is dropped for convenience.
 If $\tau=1$, we have $N_1[\ell]=\lfloor0.69]=1$ and $N_2[\ell]=\lfloor0.31\rceil=0$, respectively. If each time slot is further divided into $10$ sub-slots, i.e., $\tau=10$, then $N_1[\ell]=\lfloor6.9\rceil=7$ and $N_2[\ell]=\lfloor3.1\rceil=3$, respectively. Although such a rounding  still causes a performance gap, the gap decreases as the duration of the sub-slot decreases. Alternatively, if each time slot is divided into $100$ sub-slots, i.e., $\tau=100$,  user 1 and user 2 will be assigned 69 and 31 sub-slots, respectively, i.e., $N_{1}[\ell]=\lfloor69\rceil=69$ and $N_{2}[\ell]=\lfloor31\rceil=31$, which permits a binary solution with zero relaxation gap.  Furthermore,  since constraints (\ref{eq70}) and (\ref{eq80}) are met with equalities in the optimal solution to problem (\ref{probm25}), a binary solution for the case of multiple UAVs can be easily reconstructed by applying the above procedure.

{It is worth pointing out that such a reconstructed binary solution is always feasible for problem \eqref{probm6} with the same larger $N'$ slots, while we do not need to resolve problem \eqref{probm6} with $N'>N$ directly to avoid high computational complexity.  Thus, the complexity of our proposed approach is lower compared to that of directly solving problem  \eqref{probm6}  with $N'$ slots.} {On the other hand, the case of $\tau=1$ which directly rounds off the continuous variables to binary ones,  is a special case of the proposed scheme but at the expense of certain performance loss in general. Therefore, our proposed scheme not only ensures to obtain a feasible solution to problem  \eqref{probm6}  with any given $N$ slots, but also can achieve higher accuracy and better performance by using $N'>N$ slots yet without increasing the complexity.} {In other words, if the number of time slots $N'$ is set very large initially, then directly solving problem  \eqref{probm6}   with $N'$ will incur very high complexity. In this case, we can first formulate and solve the problem with a smaller $N=N'/\tau$ by choosing a suitable $\tau>1$ (note that $\tau$ cannot be set too large as this may render the discrete-time approximation of the UAV trajectory inaccurate),  and then use our results to construct a feasible solution to problem \eqref{probm6}  with the larger number of times slots $N'$, to achieve lower complexity.}
 \section{Numerical Results}
In this section, we provide numerical examples to demonstrate the effectiveness of the proposed algorithm. We consider a system with $K=6$ ground users that are randomly and uniformly distributed within a 2D area of $2\times 2$ km$^2$. The following results are obtained based on one random realization of the user locations as shown in Fig. \ref{single_trajectory}.  All the UAVs are assumed to fly at a fixed altitude $H=100$ m.
 The receiver noise power is assumed to be $\sigma^2= -110$ dBm. The channel power gain at the reference distance $d_0= 1$ m is set as $\rho_0=-60$ dB. The maximum transmit power and the maximum speed of UAVs are assumed as $P_{\max}=0.1$ W and $V_{\max}=50$ m/s, respectively. The threshold $\epsilon$ in Algorithm 1 is set as $10^{-4}$. The transmit power of the UAVs is initialized by the maximum transmit power, i.e., $p_m[n] = P_{\max}, \forall\, m$. Other parameters are set as $d_{\min} =100$ m and $\tau=100$.

\subsection{Singe UAV Case}

\begin{figure}[!t]
\centering
\includegraphics[width=0.4\textwidth]{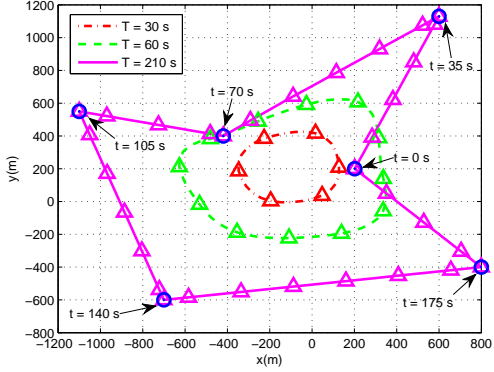}
\caption{Optimized UAV trajectories for different periods  $T$ for a single-UAV system. Each trajectory is sampled every 5 s and the sampled points are marked with `$\triangle$' by using the same colors as their corresponding trajectories. The user locations are marked by Blue circles `$\odot$'.} \label{single_trajectory}
\end{figure}
   \begin{figure}[!t]
\centering
\includegraphics[width=0.4\textwidth]{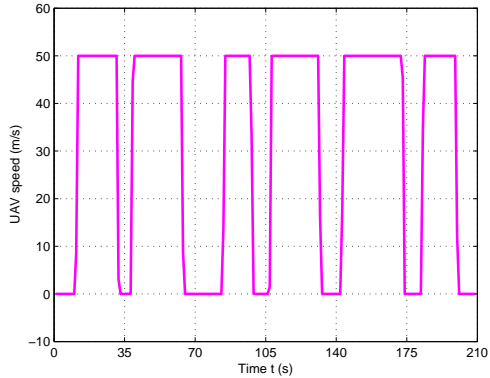}
\caption{The UAV speed versus time for $T=210$ s.} \label{speed_paper}
\end{figure}

{We first consider the special case with one single UAV, i.e., $M=1$, where there is no co-channel interference in the system. It is not difficult to see that in this case, the UAV should always transmit with its maximum power, i.e., $p[n]=P_{\max}, \forall\,n$. Then, problem (\ref{probm6}) is simplified to a joint user scheduling and UAV trajectory optimization problem that can be solved by a slight modification of Algorithm \ref{Algo:succ}.} In Fig. \ref{single_trajectory}, we illustrate the optimized trajectories obtained by the proposed Algorithm 1 under different periods $T$. It is observed that as $T$ increases, the UAV
exploits its mobility to adaptively enlarge and adjust  its trajectory to move closer to the ground users. When $T$ is sufficiently large, e.g., $T=210$ s,  the UAV is able to sequentially visit all the users and stay stationary above each of them for a certain amount of time (i.e., with a zero speed), while  the UAV trajectory becomes a closed loop with segments connecting all the points right on top of the user locations. Except the time spent on traveling between the user locations, the UAV sequentially hovers above the users so as to enjoy the best communication channels. For example, for the case of $T=210$ s, it can be observed that the sampled points on the trajectory around each user have higher densities than those far way from the users. This means that when the UAV flies close to each user, it will reduce the speed accordingly such that more information can be transmitted over a better air-to-ground channel. This phenomenon can be more directly observed from Fig. \ref{speed_paper} for the case of $T=210$ s, where the UAV speed reduces  to zero when it flies right above  each user, such as $t=35$ s. While for $T=30$ and $60$ s, the UAV always flies at the maximum speed $V_{\max}$ in order to get as close to each user as possible for shorter LoS links within each limited period $T$.

\begin{figure}[!t]
\centering
\includegraphics[width=0.4\textwidth]{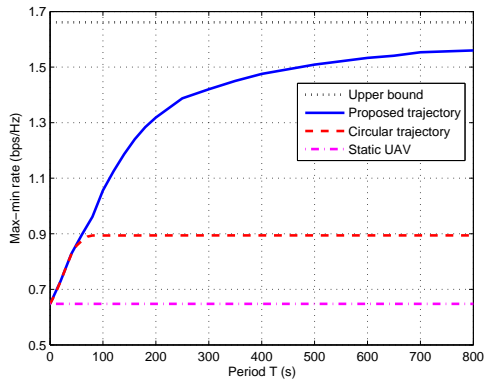}
\caption{Max-min rate  versus period $T$ for a single-UAV system with different trajectory designs.} \label{single_throughput}
\end{figure}

\begin{figure}[!t]
\centering
\includegraphics[width=0.4\textwidth]{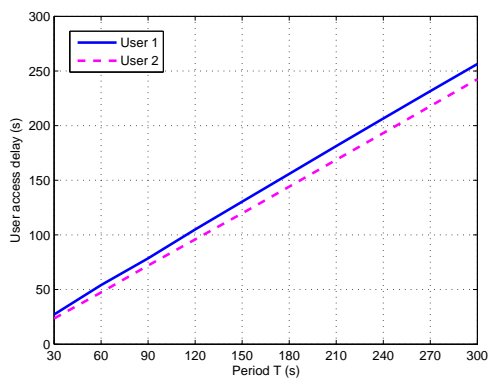}
\caption{User access delay versus period $T$ for a single-UAV system. The locations of users 1 and 2 are $[-419, 400]^T$ and $[600, 1130]^T$ in m, respectively, which are shown in Fig. \ref{single_trajectory}. } \label{user_delay}
\end{figure}

In Fig. \ref{single_throughput}, we compare the average max-min rate achieved by the following schemes: 1) Proposed trajectory, which is obtained by Algorithm 1; 2) Circular trajectory, which is obtained by the proposed  initialization scheme with $M=1$; and 3) Static UAV, where the UAV is placed at the geometric center of the user positions and remains static during the whole period $T$. For all the three schemes, the user scheduling is optimized by Algorithm 1 with given trajectory. It is observed from Fig. \ref{single_throughput} that the max-min rate of the static UAV is independent of $T$ since without mobility, the channel links between the UAV and users are time-invariant. In contrast, for the proposed trajectory and the circular trajectory schemes, the max-min rate increases with $T$ and eventually becomes saturated when $T$ is sufficiently large. This is expected since with the UAV mobility,  a larger $T$ provides the UAV more time to fly closer to the users to be served, which thus improves the max-min rate. {In addition, when $T$ and/or $V_{\max}$ is sufficiently large such that the UAV's travelling time between users is negligible, each ground user is sequentially served with equal time duration when the UAV is directly on top of it. In this case,  the max-min rate for each user can be obtained as }
\begin{align}\label{upperbound}
R^{\rm ub} = \frac{1}{K}  \log_2\left( 1 + {\frac{P\rho_0}{H^2\sigma^2}}\right)= 1.6612\  \text{bps/Hz}.
\end{align}
{It is worth pointing out that since the travelling time in practice is always not negligible  for any finite UAV speed,  the maximum objective value of problem \eqref{probm6} is strictly upper-bounded by the rate  in \eqref{upperbound}. As the obtained trajectory by our proposed algorithm is able to move the UAV to be above of each user,  the asymptotic optimality of the proposed algorithm can be demonstrated with increasing $T$, which can be seen in Fig. \ref{single_throughput}. In Fig. \ref{user_delay}, we plot the access delay for two of the users versus the period $T$ based on the optimized user scheduling variables.
One can observe that as $T$ increases, the user access delay also increases, which implies that each user needs to wait for a longer time to be scheduled for communication with the UAV. Based on Figs. \ref{single_throughput} and \ref{user_delay}, the fundamental delay-throughput tradeoff is demonstrated.  }

By comparing the performance of the proposed trajectory with that of the circular trajectory in Fig. \ref{single_throughput},  the advantage of fully exploiting the trajectory design  is also demonstrated. Since the circular trajectory restricts the UAV to fly along a circle, the users that are not around the circle suffer from worse channels.  As a result,  more time needs to be assigned to those users, which poses the bottleneck for the achievable max-min throughput.
While for the proposed trajectory with a sufficiently large period $T$, the UAV is able to fly closer to or even stays stationary above all users to serve them with better channels. Therefore, the max-min throughput is improved, but at the cost of longer access delay on average for the users.

   \begin{figure}[!t]
\centering
\includegraphics[width=0.4\textwidth]{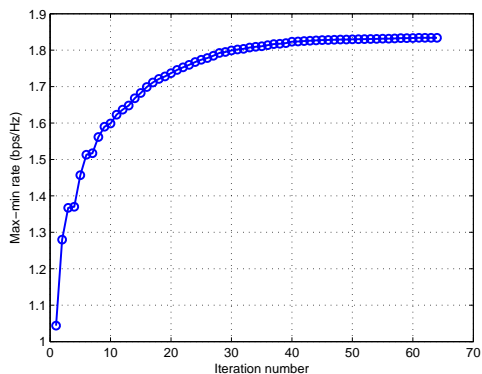}
\caption{Convergence behaviour of the proposed Algorithm 1.} \label{convergence}
\end{figure}

\subsection{Multi-UAVs Case}

Next, we study the max-min throughput of the  multi-UAV network.  Before the performance comparison, we show the convergence behaviour of the proposed Algorithm 1  in Fig. \ref{convergence} for the case of two UAVs under $T=90$ s.  It can be observed from the figure that the max-min rate achieved by the proposed algorithm increases quickly with the number of iterations and the algorithm converges in about 40 iterations.

   \begin{figure}[!t]
\centering
\includegraphics[width=0.4\textwidth]{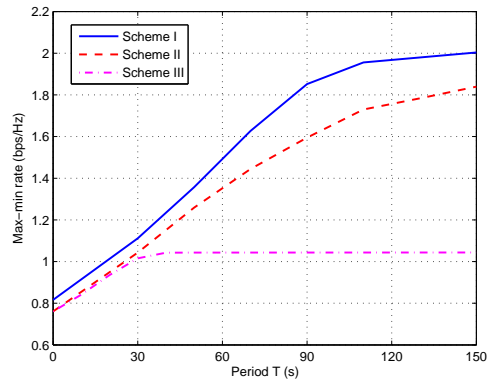}
\caption{Max-min rate versus period $T$ for a two-UAV system with different optimization schemes.} \label{twoUAV_T}
\end{figure}

In order to show the performance gain brought by the optimization of the different design variables  in Algorithm 1,
 in Fig. \ref{twoUAV_T}, we compare the following three schemes for a two-UAV network, namely,  1) Scheme I: All variables are jointly optimized as in Algorithm 1; 2) Scheme II: Jointly optimized  user scheduling and association as well as UAV trajectory but with full transmit power (i.e., no transmit power control); and 3) Scheme III: Optimized user scheduling and association but with simple  circular trajectory and full transmit power of UAVs. Several important observations can be made from Fig. \ref{twoUAV_T}. First, as expected,  the max-min rates  of all the three schemes increase as the period $T$ becomes large. 
Second,  the performance gap between Scheme II and Scheme III demonstrates the throughput gain brought by the proposed trajectory design even without transmit power control applied, and the performance gap between the two schemes increases with increasing $T$.  This is because with larger $T$, the optimization of  UAVs' trajectories becomes more crucial for both achieving better direct links and avoiding severe co-channel interference links, especially when there is no transmit power control applied, whereas restricting the UAVs flying along circles limits the potential of UAV mobility.
  Second, by comparing Scheme I and Scheme II, the additional gain of power control is also demonstrated. When the power control can be optimized, it also provides more flexibility for designing UAVs' trajectories, which helps achieve better user rates.
Last but not the least, by comparing Scheme I and its counterpart for the case of a single UAV in Fig. \ref{single_throughput}, it is observed that the user access delay is significantly reduced by employing two UAVs to serve users jointly. For example, to achieve the same average max-min rate about $1.60$ bps/Hz, a single-UAV system requires more than $T=800$ s as shown in Fig. \ref{single_throughput}, whereas this value dramatically reduces to about $T=70$ s for a two-UAV system, both applying the proposed Algorithm 1.
 Such a performance gain is mainly attributed to two facts. On one hand,  the spectrum efficiency is improved by allowing concurrent transmissions of the  two UAVs with the same power budget. In fact, this can be directly observed by comparing  the upper bound of the max-min rate for a single-UAV system which  is $1.6612$ bps/Hz given in  (\ref{upperbound}) with the achievable max-min rate of the two-UAV system which is more than  $2.00$ bps/Hz as shown in Fig. \ref{twoUAV_T}.
  On the other hand, the traveling time of each UAV over its served  ground users is reduced and the average air-to-ground channels are also improved when the number of UAVs increases,  which saves more time for them to stay above  each user to maintain the best LoS channels. In summary, the above observations demonstrate the effectiveness of employing multiple UAVs  for improving the user throughput and/or reducing the access delay, which thus improves the fundamental throughput-access delay tradeoff.


\begin{figure}[!t]
\centering
\subfigure[Optimized UAV trajectories without power control.]{\includegraphics[width=0.4\textwidth]{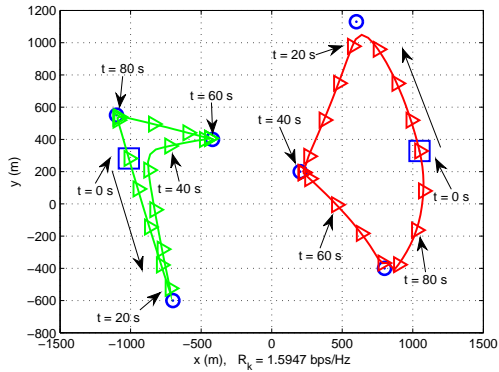}} 
\subfigure[Optimized  UAV trajectories with power control.]{\includegraphics[width=0.4\textwidth]{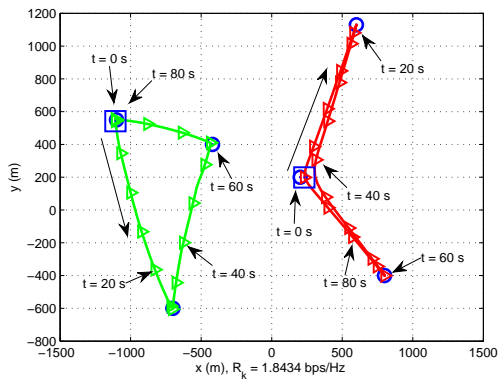}}
\caption{ Trajectory comparison for a two-UAV system when $T=90$ s. The initial locations of trajectories are marked with blue square `$\Box$'. Black arrows represent the directions of the trajectories. Each trajectory is sampled every 5 s and the sampling points are marked with `$\triangle$'s by using the same colors as their corresponding trajectories. } \label{twouav_trj}\vspace{-0.5cm}
\end{figure}

   \begin{figure}[!t]
\centering
\includegraphics[width=0.4\textwidth]{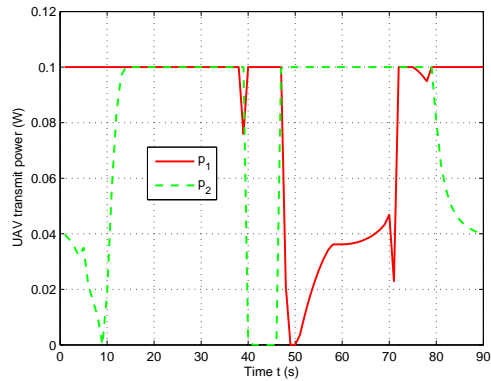}
\caption{UAV transmit power versus time for a two-UAV system.} \label{transmit_power}
\end{figure}

In Fig. \ref{twouav_trj}, we compare the optimized UAV trajectories obtained by Schemes I and II with the period $T=90$ s. {It can be observed from Fig. \ref{twouav_trj} (a)  that for Scheme II without power control, i.e., when the maximum transmit power is used by both UAVs, the trajectories of the two UAVs tend to keep away from each other as far as possible to avoid co-channel interference. However, at some pair of UAV locations, this is realized at the cost of sacrificing favourable direct  communication links, especially when they have to serve two users that are close to each other. As a result, the advantage of trajectory design is compromised so as to trade off between the direct channel and the co-channel interference. In contrast, in Fig \ref{twouav_trj} (b) for Scheme I when the transmit power is also optimized,  the two  UAVs can reduce the interference by properly adjusting the transmit power when they get close to each other to serve nearby users. As such,  strong direct links and weak co-channel interference can be achieved at the same time, which helps unlock the potential benefit brought by the trajectory design and thereby achieves a larger max-min rate ($R_k=1.8434$ bps/Hz, $\forall\, k$, with Scheme I versus $R_k=1.5947$ bps/Hz, $\forall\, k$, with Scheme II).}
{The corresponding UAV transmit power versus time is plotted in Fig. \ref{transmit_power}. First, it can be observed that  at any time instant, there is always one UAV that transmits with the maximum power.
 Second, when two UAVs are far away from each other, both of them tend to transmit with the maximum power so as to improve the spectrum efficiency, e.g., from $t=10$ s to $t=20$ s where two UAVs flight towards the opposite directions. In contrast, when the two UAVs are getting very close to each other, one UAV will reduce the transmit power to zero to avoid severe interference, e.g., from $t=40$ s to $t=45$ s where the two UAVs are serving the two nearby users in the center. Therefore, without power control, the communication interference can only be mitigated by adjusting the UAV trajectory, while a joint power control and trajectory  design provides more  flexibility to mitigate the co-channel interference  and thus achieves a higher max-min rate. }

   \begin{figure}[!t]
\centering
\includegraphics[width=0.4\textwidth]{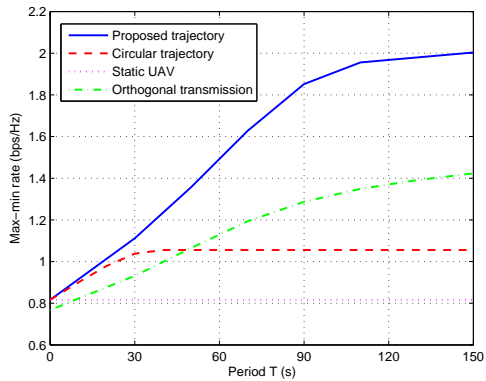}
\caption{Max-min rate  versus period $T$ for a two-UAV system with different trajectory designs and the orthogonal transmission.} \label{twoUAV_throughput}
\end{figure}

{In Fig. \ref{twoUAV_throughput}, we compare the average max-min rate achieved by the three trajectory designs in a two-UAV system similar to those  in Fig. \ref{single_throughput} for the single-UAV case, i.e., 1) Proposed trajectory; 2) Circular trajectory, which is obtained by the proposed  initialization scheme with $M=2$; and 3) Static UAV, where each  UAV  $m$ is placed at $\mathbf{c}^m_{\rm trj}$ as in the initialization scheme and remains static for the entire $T$.
 For all the three schemes, both the user scheduling and association as well as power control are optimized by Algorithm 1 with given corresponding trajectory.} {In addition, an orthogonal UAV transmission scheme is also adopted for comparison. Specifically, the multiple UAVs take turns to transmit information to their served ground users over orthogonal time slots, thus the system is interference-free. This is achieved by imposing the following constraints\footnote{ {For convenience, we select the value of $N$ such that $\frac{N}{M}$ is an integer for a given $M$.}},
\begin{align}
\kern -2mm &\sum_{k=1}^{K}\alpha_{k,m}[M(\ell-1)+m]\leq 1, \forall\,m, \ell=1,\cdot\cdot\cdot,\frac{N}{M}, \label{eq220}\\
\kern -2mm &\sum_{k=1}^{K}\alpha_{k,j}[M(\ell-1)+m]= 0, \forall\,j\neq m, \ell=1,\cdot\cdot\cdot,\frac{N}{M}, \label{eq221}
 \end{align}}
{\kern -1.6mm which  guarantee that in each time slot, only one UAV is allowed to transmit.  Accordingly, the achievable rate of user $k$ can be  expressed as
{\myfont \begin{align}
 R^{II}_k=\frac{1}{N}\sum_{n=1}^{N}\sum_{m=1}^{M}\alpha_{k,m}[n]  \log_2\left( 1 + {\frac{p_m[n]\rho_0}{(H^2+||\mathbf{q}_m[n]-\mathbf{w}_k||^2)\sigma^2}}\right).
 \end{align}}
Since the above case  is a special case of problem (\ref{probm6}), the corresponding problem  can be solved similarly by Algorithm 1.}
  As can be seen, the max-min rate of the static-UAV case is still regardless of the period $T$ due to the time-invariant air-to-ground channels. In contrast, by exploiting the UAV mobility,  the max-min rates achieved by the other two trajectory designs  are non-decreasing with $T$, which further demonstrates the fundamental throughput-access delay tradeoff. Compared to Fig. \ref{single_throughput} with a single UAV, it can also be observed that such a tradeoff has been significantly improved (i.e., higher max-min rate is achieved with the same given $T$)  by employing more than one  UAVs. In addition, compared to the orthogonal transmission scheme, the spectrum sharing gain by the two UAVs  is also demonstrated.

\section{Conclusions}
In this paper, we have investigated a new type of multi-UAV enabled wireless networks. Specifically,  the user scheduling and association, UAV trajectories, and transmit power are jointly optimized with the objective of maximizing the minimum average rate among all users.  By means of the block coordinate descent and  the successive convex optimization techniques, an efficient iterative algorithm has been proposed, which is guaranteed to converge.
 Numerical results demonstrate that the UAV mobility provides the benefit of achieving better  air-to-ground channels as well as additional flexibility for interference mitigation, and thereby improves the system throughput, compared to the conventional case with static BSs. Furthermore, the proposed trajectory design significantly outperforms the simple circular trajectory. The interesting throughput-access delay tradeoff is also shown for multi-UAV enabled communications.

 {Although we focus on the downlink communication scenario from the UAVs to ground users, the problem for the uplink communication scenario from ground users to the UAVs can be pursued by following a similar approach via optimizing the UAV trajectory alternately with the joint optimization of user scheduling and power control. However, how to integrate the solution of the joint optimization of user scheduling and power control into the framework of the block coordinate descent method to guarantee the convergence is challenging and needs further investigation.}
In addition, there are still many other interesting research directions that could be pursued in future work by extending the results of this paper, including e.g.  1) Co-existence design of a network with both aerial and ground BSs; 2) 3-D UAV trajectory design with both altitude and horizontal position optimization; and 3) Energy-efficient UAV trajectory design for the general multi-UAV and/or  multi-user scenario by taking into account the UAV movement energy consumption \cite{zeng2016energy}.  

\bibliographystyle{IEEEtran}
\bibliography{IEEEabrv,mybib}

\end{document}